\documentstyle[12pt,a4wide,epsf]{article} 
\begin{document} 
 
\begin{flushright} 
\today \\ 
arxiv \\ 
\end{flushright} 
\vskip 1cm 
\begin{center} 
{\large \bf Small Scale Structure Formation in Chameleon Cosmology} 
\end{center} 
\vspace*{5mm} \noindent 
 
\centerline{Ph.~Brax\footnote{brax@spht.saclay.cea.fr}${}^{a}$, 
C. van de Bruck\footnote{C.vandeBruck@sheffield.ac.uk}${}^{b}$ 
A. C. Davis\footnote{a.c.davis@damtp.cam .ac.uk}${}^{c}$ 
and A. M. Green\footnote{a.m.green@sheffield.ac.uk}${}^{d}$ 
} 
\vskip 0.5cm \centerline{${a}$) \em Service de Physique 
Th\'eorique} \centerline{\em CEA/DSM/SPhT, Unit\'e de recherche 
associ\'ee au CNRS,} \centerline{\em CEA-Saclay F-91191 Gif/Yvette 
cedex, France.} \vskip 0.5cm 
\vskip 0.5cm \centerline{${b}$) \em Department of Applied Mathematics} 
\centerline{\em The University of Sheffield} 
\centerline{\em Hounsfield Road, Sheffield S3 7RH, United Kingdom} \vskip 0.5cm 
 
\vskip 0.5cm \centerline{${c}$) \em Department of Applied Mathematics and Theoretical Physics} 
\centerline{\em Centre for Mathematical Sciences} 
\centerline{\em Cambridge CB2 0WA, United Kingdom} \vskip 0.5cm 
\vskip 0.5cm \centerline{${d}$) \em Department of Physics and Astronomy} 
\centerline{\em The University of Sheffield} 
\centerline{\em Hounsfield Road, Sheffield S3 7RH, United Kingdom} \vskip 0.5cm 
 
\begin{abstract} 

Chameleon fields are scalar fields whose mass depends on the
ambient matter density. 
We investigate the
effects of these fields on the growth of density perturbations 
on sub-galactic scales and the
formation of the first dark matter halos.  
Density perturbations on comoving scales $R < 1 {\rm pc}$ go
non--linear and collapse to form structure much earlier than in
standard $\Lambda$CDM cosmology. The resulting
mini-halos are hence more dense and
resilient to disruption. We therefore expect (provided
that the density perturbations on these scales have not been 
erased by damping processes) that the dark matter
distribution on small scales would be more clumpy in chameleon
cosmology than in the $\Lambda$CDM model. 

\end{abstract}

\newpage 
 
\section{Introduction} 
 
Cosmological observations~\cite{cosmo1,cosmo2} indicate that more 
than two thirds of the energy density of the Universe is in a 
component with negative pressure. Candidates for this missing 
energy, which is causing the Universe to accelerate, include a 
cosmological constant and scalar field models with equation of state 
$w \neq 1$, often referred to as quintessence~\cite{quint}. In these 
models, the scalar field is either decoupled from Cold Dark Matter 
(CDM) or couples to CDM but not to the baryons (coupled 
quintessence) 
 
In order to generate the present day acceleration the scalar field in
these models must be evolving slowly and hence have a tiny mass, of
order the present day Hubble constant, $H_{0} \sim 10^{-33} {\rm
eV}$. Since the mass of a quintessence field is very small, it can
give rise to a new long-range force. Such a force has not been
observed and consequently the coupling of the quintessence field to
matter must be very small. Unfortunately, in effective theories
derived from string theory nearly massless fields couple to matter
with gravitational strength and would produce unacceptably large
violations of the equivalence principle.  Khoury and Weltman have
proposed a scenario where a scalar field with gravitational strength
coupling to matter can evolve on a Hubble timescale and generate the
present day acceleration while evading all existing tests of gravity
\cite{cham1}. The key feature of this scenario is that the scalar
field, which is dubbed the chameleon, has a mass which depends on the
local background matter density. On Earth where the density is high,
the Compton wavelength of the field is sufficiently small to satisfy
all current tests of gravity, while on cosmological scales, where the
density is tiny, the field has a much smaller mass and can drive the
present day acceleration \cite{cham2}. The field, however, is heavier
than in standard quintessence models ($m_{\rm cham} \gg H$).  In the
solar system, where the density is many orders of magnitude smaller
than on Earth, the chameleon is essentially a free field and mediates
a long range force which could be detected by upcoming satellite
experiments~\cite{exp}.
 
The cosmological history of the chameleon field was studied in 
Ref.~\cite{cham2}.  It was shown that there is an attractor solution, 
analogous to the tracker solution in quintessence models, where the 
chameleon quickly settles into the minimum of its effective potential, 
and for a broad class of potentials and initial conditions the 
chameleon can satisfy all observational constraints. The evolution of 
the density perturbations was also investigated. It was shown that 
perturbations on scales smaller than the scale of the chameleon feel a 
larger effective Newton's constant which causes them to grow more 
rapidly. The length scale of the chameleon (${\cal O} (100 \, {\rm 
pc})$ at present) is somewhat smaller than the scales probed by large 
scale structure observations. There has, however, recently been much 
interest in the properties of the first generation of dark matter 
structures to form in the Universe, and the possibility that they may 
leave an observable imprint in the present day dark matter 
distribution~\cite{hss,bde,wimp1,dms,wimp2}. 
 
In this paper we examine the effects of the chameleon on the density 
perturbations on sub-galactic scales and the properties (formation 
epoch and over-density) of the first gravitationally bound structures 
to form. In section~\ref{chamdyn} we review the necessary aspects of 
the chameleon and its dynamics from Ref.~\cite{cham2}. In 
section~\ref{pert} we extend the calculations of the evolution of 
density perturbations in Refs.~\cite{wimp1,wimp2} to include the 
effects of the chameleon. Finally in section~\ref{sscham} we examine 
the effect of the chameleon on the formation of small scale structure.

\section{The chameleon and its dynamics} 
\label{chamdyn} 
 
The action describing the chameleon field, $\chi$, matter and gravity has the 
general form 
\begin{equation} 
S= \int {\rm d}^4 x \sqrt{-g} \left[ \frac{M_{\rm Pl}^2}{2} {\cal R} 
  - \frac{1}{2} (\partial \chi)^2 - V(\chi) \right] - \int {\rm d}^4 x 
      {\cal L}_{\rm m}(\chi_{\rm m}^{({\rm i})} g_{\mu \nu}^{({\rm i})}) \,, 
\end{equation} 
where $M_{\rm Pl} \equiv (8 \pi G)^{-1/2}$ is the reduced Planck 
mass and $\chi_{\rm m}^{({\rm i})}$ are the various matter fields. 
The metrics governing the excitations of the matter fields are 
related to the Einstein frame metric $g_{\mu \nu}$ via the conformal 
rescaling $g_{\mu \nu}^{({\rm i})} = \exp{(2 \beta_{i} \phi/ M_{\rm 
Pl})} g_{\mu \nu} $ where the $\beta_{\rm i}$ are dimensionless 
quantities of order unity~\cite{beta}. Notice that the scalar field 
couples to all matter species including the baryons. As an example,  
this is the case for the radion field describing the interbrane distance  
in brane world models based on the Randall--Sundrum model \cite{rs1} in which branes  
are nearby, for which $\beta_i\equiv \beta=1/\sqrt 6$. 
 
By varying the action with respect to $\chi$ it can be 
shown~\cite{beta,peccei,cham1,cham2} that the dynamics of $\chi$ are governed by 
the effective potential 
\begin{equation} 
V_{\rm eff}(\chi) = V(\chi) + \Sigma_{i} \rho_{i} 
       \exp{(\beta_{\rm i} \chi/ M_{\rm Pl})} \,, 
\end{equation} 
where the matter density $\rho_{\rm i}$ is defined as $\rho_{\rm i} \equiv 
- g_{({\rm i})}^{\mu \nu} T_{\mu \nu}^{({\rm i})} 
\exp{(3 \beta_{\rm i} \chi/ M_{\rm Pl})}$ so that it is independent of $\chi$ 
and conserved in the Einstein frame. If $V({\chi})$ decreases monotonically 
with increasing $\chi$ and $\beta_{\rm i} > 0$, this potential has a minimum, 
$\chi_{\rm min}$, which increases with decreasing $\rho_{\rm i}$. 
The mass of small fluctuations about $\chi_{\rm min}$ increases with 
increasing $\chi_{\rm min}$ so that the chameleon can evade local 
tests of the equivalence principle and fifth forces, due to the high local 
density. 
 
Fiducial potentials of the form 
\begin{equation} 
V(\chi) = M^{4} \exp{(M^{n}/\chi^{n})} \,, 
\end{equation} 
with $M=10^{-3} {\rm eV}$ so as to produce the observed present day dark 
energy density (and also satisfy local tests of general relativity) were 
studied in Ref.~\cite{cham2}. 
Assuming for simplicity a single matter component with density 
$\rho_{\rm m}$ and coupling $\beta$, the field value at the minimum 
of the effective potential satisfies 
\begin{equation} 
\label{phimin} 
\left( \frac{M}{\chi_{\rm min}(t)} \right)^{n+1} = 
      \frac{\beta}{n} \frac{M}{M_{\rm Pl}} \frac{ \rho_{\rm m} 
      \exp{(\beta \chi_{\rm min}(t) /M_{\rm Pl})}}{V(\chi_{\rm min}(t))} \,. 
\end{equation} 
It was found that for a wide range of initial conditions the chameleon 
field reaches the attractor solution with $\chi(t)=\chi_{\rm min}(t)$ 
before big bang nucleosynthesis and has a cosmological evolution in 
accordance with all observational constraints. 
 
The presence of a chameleon field can effect the growth of structure 
in particular on small scales. In the following, we will consider 
the modifications to the growth factor due to the gravitational 
effects induced by the chameleon.

\section{Perturbation evolution} 
\label{pert} 
We will follow Refs.~\cite{wimp1,wimp2} and work in the longitudinal 
gauge, we will however use the notation of Ma and Bertschinger~\cite{mb}. 
The perturbed line element reads 
\begin{equation} 
ds^2 = a^2(\eta) \left[-\left(1+2\Psi\right)d\eta^2 + 
         \left(1 - 2\phi\right)g_{ij}dx^i dx^j \right] \,. 
\end{equation} 
The equations of motion for the CDM density 
contrast $\delta_{\rm c}$ and the 
divergence of the CDM velocity field $\Theta_{\rm c}$ 
can be obtained 
from the energy-momentum conservation equation, which contains an additional 
term due to the exchange of energy with the chameleon 
\begin{equation} 
T^{\mu\nu}_{(i)~;\mu} = \beta_{(i)} (\partial^\nu \chi) T_{(i)} \,. 
\end{equation} 
Here, $i$ stands for the component $i$ and $T = T^{\mu}_{~\mu}$. The 
equations of 
motion are then given by (the dot represents the derivative with 
respect to $\eta$ and ${\cal H} 
\equiv ({\rm d} a/ {\rm d} \eta)/a$) 
\begin{eqnarray} 
\label{e1} 
\dot{\delta_{\rm c}} &=& -\Theta_{\rm c} + 3\dot{\phi} + \beta(\delta\chi)^. 
    \,, \\ 
\dot{\Theta_{\rm c}} &=& - ( {\cal H} + \beta \dot{\chi})\Theta_{\rm c} + 
     k^2\left(\Psi + \beta\delta\chi\right) \,. 
\end{eqnarray} 
The perturbed Klein-Gordon equation for the chameleon field $\chi$ 
is given by 
\begin{equation} 
\label{klein} 
(\delta\chi)^{..} + 2 {\cal H}(\delta\chi)^. + \left( k^2 + a^2 
     \frac{\partial^2 V}{\partial \chi^2} \right)\delta\chi 
  + 2\Psi\left( \frac{\partial V}{\partial \chi} + \beta\rho_{\rm c} 
     \right)a^2 
     - 4\dot\Psi\dot\chi =- \beta\rho_{\rm c} \delta_{\rm c} a^2 \,, 
\end{equation} 
and we will also need one of the components of the first-order perturbed 
Einstein equation (Poisson's equation in the sub-horizon limit) 
\begin{equation} 
\label{poisson} 
k^2 \phi + 3 {\cal H} ( \dot{\phi} + {\cal H} \psi) = 4 \pi G a^{2} 
       \delta T^{0}_{0} \,. 
\end{equation}

From very early times onwards (before nucleosynthesis), the mass of the chameleon field is much greater 
than the Hubble expansion and the field sits in the minimum of the 
effective potential. Consequently the interaction scale of the 
chameleon field is always much smaller than the horizon $H^{-1}$, and 
the evolution of perturbations on super-horizon scales is unaffected 
by the chameleon. Furthermore the chameleon does not couple to 
radiation, since it is a traceless fluid, and the evolution of 
perturbations deep within the radiation dominated epoch is also as in 
standard cosmology~\cite{mb,wimp2}. Once $\delta_{\rm c} \rho_{\rm c} 
\gg \delta_{\rm r} \rho_{\rm r}$ (which for sub-galactic scales 
happens prior to matter-radiation equality), however, the dark matter 
terms dominate as the source in the Poisson equation, 
eq.~(\ref{poisson}), and the coupling of the chameleon to the matter density 
is now important. In particular, perturbations in matter will influence 
perturbations in the chameleon field and vice versa. 
 
On the sub-horizon scales we are interested in we can neglect the 
oscillations in the perturbations in the chameleon field, and $\dot\chi$ is also small as 
the field evolves along the minimum of the effective potential 
\cite{cham2}. Following Ref.~\cite{wimp2} we also neglect anisotropic 
stress, so that $\phi=\psi$, and baryon anisotropies (but not the 
baryon density).  The latter assumption is valid on small scales ($k> 
k_{\rm b} \sim 10^{-3} {\rm pc}^{-1}$) for $ z> z_{\rm b} \sim 150$ 
as prior to this residual electrons allow transfer of energy between the photon 
and baryon fluids and thermal pressure prevents the baryon 
perturbations from growing~\cite{baryons}. 
Following Ref.~\cite{cham2,wimp2}, eqs.~(\ref{e1}-\ref{poisson}) 
can be combined to give the following equation for the evolution of 
the cold dark matter density contrast: 
\begin{equation} 
\label{del1} 
\ddot\delta_c = - {\cal H} \dot\delta_c + 
     \frac{3}{2}\frac{\rho_c}{\rho_c+\rho_\gamma}\left[ 
1 + \frac{2\beta^2}{1+\frac{a^2 V^{''}}{k^2}} \right]\delta_c, 
\end{equation} 
with $V^{''} = \partial^2 V/\partial \chi^2$. The 
effects of the chameleon manifest themselves in the second term in the 
square brackets. The chameleon field operates on  length scales 
smaller than 
\begin{equation} 
\label{lambdacham} 
\lambda_{\rm cham}(t) \equiv \frac{1}{\sqrt{V''}} \,, 
\end{equation} 
or equivalently for comoving wavenumbers larger than 
\begin{equation} 
\label{kcham} 
k_{\rm cham}(t) = \frac{a}{\lambda_{\rm cham}(t)} \,. 
\end{equation} 
For $k \ll k_{\rm cham}(t)$ the terms in the square 
bracket in eq.~(\ref{del1}) above are well 
approximated by 1, and the CDM density contrast evolves 
as in standard cosmology. For $k \gg k_{\rm cham}(t)$, they are 
well approximated by $(1+2\beta^2)$ i.e. on these scales the 
growth of perturbations is governed by an effective gravitational 
constant given by $G(1+2\beta^2)$. 
 
For $k \gg k_{\rm cham}$, Eq.~(\ref{del1}) can be re-written in 
terms of $y=a/a_{\rm eq}$ to give 
\begin{equation} 
\label{del2} 
y(y+1)\delta_{\rm c}^{''} + \left( 1 + \frac{3}{2}y \right) \delta_{\rm c}^{'} 
     = \frac{3}{2}\left(1+2\beta^2\right) 
\left(1-f_b\right)\delta_{\rm c}, 
\end{equation} 
where ${}^{'}= {\rm d} /{\rm d} y$ and we have 
introduced the baryon fraction $f_b \equiv \Omega_{\rm 
b}/ \Omega_{\rm m}$. The solution 
to this equation is a superposition of Legendre functions of 
first and second kind of order $\nu$: 
 
\begin{equation} 
\label{legendre} 
\delta_c(k,y) = B_1(k)P_\nu\left(\sqrt{1+y}\right) + 
     B_2(k)Q_\nu\left(\sqrt{1+y}\right). 
\end{equation} 
 
This is the same as in the $\Lambda$CDM case \cite{wimp1,wimp2}, but 
the degree $\nu$ of the Legendre functions in the standard case is given by 
\begin{equation} 
\label{nuGR} 
\nu_{\rm GR} = \frac{-1 + \sqrt{1 + 24(1-f_{\rm b})}}{2} \,, 
\end{equation} 
whereas we have, for $k \gg k_{\rm cham}(t)$, 
\begin{equation} 
\label{nucham} 
\nu_{\rm cham} = \frac{-1 + \sqrt{1 + 24(1+ 2 \beta^2)(1-f_{\rm b})}}{2} \,. 
\end{equation} 
For the best fit WMAP $\Lambda$CDM model $f_b = 0.17$~\cite{cosmo1} 
so that $\nu_{\rm GR} = 1.8$, whereas for $\beta=1$ $\nu_{\rm 
cham}=3.4$. For $z < z_{\rm b} \sim 150$ the baryons follow the CDM 
(which is equivalent to setting $f_{\rm b}=0$ in 
eqs.~(\ref{del2})-(\ref{nucham})) and $\nu_{{\rm GR}}=2$ for the 
standard cosmology and $\nu_{{\rm cham}}=3.8$ for the chameleon with 
$\beta=1$ (as found in Ref.~\cite{cham2}). For $y \gg 1$, 
equivalently $z \ll z_{\rm eq}$, $\delta_{\rm c}(y)$ grows as $ 
a^{\nu/2}$. The increase in the growth rate due to the chameleon is 
large compared with the suppression in growth due to baryons which 
occurs for $z_{\rm b} < z < z_{\rm eq}$. We therefore now neglect 
the baryons and set $f_{\rm b}=0$ in eqns.~(\ref{nuGR}) and 
(\ref{nucham}) above. 
 
The full asymptotic late time solution is 
~\cite{wimp1,wimp2} 
\begin{equation} 
\delta_{\rm c}(y) = 6\zeta_0 c(\nu)y^{\nu/2} 
     \left[\ln\left(\frac{k}{k_{\rm eq}}\right) + b(\nu) \right] \,. 
\end{equation} 
where $\zeta_{0}$ is the superhorizon limit of the curvature 
perturbation on uniform density hypersurfaces and  the constants 
$c(\nu)$ and $b(\nu)$ are found by matching the early time 
radiation domination ($y\ll 1$) expansion of eq.~(\ref{legendre}) to 
the sub-horizon limit of the general radiation domination solution~\cite{wimp1,wimp2} 
\begin{equation} 
c(\nu) = \frac{\Gamma(1+2\nu)}{2^\nu\Gamma^2(1+\nu)}, 
\end{equation} 
and 
\begin{equation} 
b(\nu) = \frac{1}{2}\ln\left(\frac{2^5}{3}\right) - \gamma_E - 
  \frac{1}{2} - \frac{2}{\nu} - \frac{2\Gamma'(\nu)}{\Gamma(\nu)}. 
\end{equation} 
where $\Gamma'(\nu)$ is the derivative of $\Gamma(\nu)$ with respect to $\nu$. 
For $\nu_{\rm GR}=2$, $c=1.5$ and $b=-1.7$, while for 
$\nu_{\rm cham}=3.8$, $c=3.9$ and $b=-2.8$. 
 
We now calculate the evolution of $k_{\rm cham}(t)$ with time, and consequently 
the scales on which the chameleon effects the growth of the density contrast. 
During matter domination $V(\chi_{\rm min}) \sim {\rm const} $ and 
$\rho_{\rm m} 
 \exp{(\beta \chi_{\rm min} /M_{\rm Pl})} \propto a^{-3}$ 
so that, using eq.~(\ref{phimin}), 
\begin{equation} 
\chi_{\rm min} (t) = \chi_{\rm min} (t_{0}) \left(1+z \right)^{-3/(n+1)} \,. 
\end{equation} 
At present $\rho_{\rm m} 
 \exp{(\beta \phi_{\rm min} /M_{\rm Pl})} \sim V(\phi_{\rm min}) 
\sim M^4$ so that~\cite{cham2} 
\begin{equation} 
\chi_{\rm min}(t_{0}) = \left( \frac{n}{\beta} \right)^{1/(n+1)} 
           \left( \frac{M}{M_{\rm Pl}} \right)^{n/(n+1)} M_{\rm Pl} \,, 
\end{equation} 
Using eqs.~(\ref{lambdacham}) and (\ref{kcham}) 
the scale of the chameleon field varies as 
\begin{eqnarray} 
\lambda_{\rm cham}(t)& = &\lambda_{\rm cham}(t_{0}) (1+z)^{[-3(n+2)/2(n+1)]} 
       \,, \\ 
k_{\rm cham}(t) &=& k_{\rm cham}(t_{0}) (1+z)^{(n+4)/2(n+1)} 
     = \frac{(1+z)^{(n+4)/2(n+1)}}{\lambda_{\rm cham}(t_{0})} \,. 
\end{eqnarray} 
where 
\begin{equation} 
\lambda_{\rm cham}(t_{0}) = \frac{1}{\sqrt{n(n+1)}} 
          \frac{1}{M}   \left( \frac{n}{\beta} \right)^{(n+2)/2(n+1)} 
   \left( \frac{M_{\rm Pl}}{M} \right)^{(n+2)/2(n+1)} \,, 
\end{equation} 
The characteristic wavenumber
$k_{\rm cham}(t)$ decreases with decreasing red-shift. 
At $z_{\rm eq}$ 
\begin{equation} 
k_{\rm cham}(z_{\rm eq}) = \frac{(1+z_{\rm eq})^{(n+4)/2(n+1)}} 
                    {\lambda_{\rm cham}(t_{0})} \,, 
\end{equation} 
where $(1+z_{\rm eq}) = 24000 \Omega_{\rm m} h^{2} \approx 3700$. 
For fiducial parameters $n=\beta=1$ and $M=10^{-3} {\rm eV}$, 
$\lambda_{\rm cham}(t_{0})= 250 \, {\rm pc}$ 
and $k_{\rm cham}(t_{\rm eq}) = 120 \, {\rm pc}^{-1} $. 
 
For scales $k<k_{\rm cham}(z_{\rm eq})$ the density 
contrast growth law is only 
modified once $k>k_{\rm cham}(t)$. 
We denote the redshift at which 
this happens by $z_{\rm mod}$, which is given by 
\begin{equation} 
\label{grow} 
(1+z_{\rm mod}) \approx 
   \left( \lambda_{\rm cham}(t_{0}) k \right)^{2(n+1)/(n+4)} 
       \,. 
\end{equation} 
For $z \ll z_{\rm eq}$ density perturbations grow as $\delta \propto 
a^{\nu/2}$, where for standard general relativity and $k \ll k_{\rm cham}(t)$, 
$\nu=\nu_{\rm GR}$ while for $k \gg 
k_{\rm cham}(t)$, $\nu=\nu_{\rm cham}$. The, scale dependent, additional 
growth in the linear density 
perturbation due to the chameleon is 
given by 
\begin{equation} 
\label{chamgrow1} 
\frac{\delta_{\rm cham}(k,z)}{\delta_{\rm GR}(k,z)} = 
  \left( \frac{1+z_{\rm mod}(k)}{1+z} 
   \right)^{(\nu_{\rm cham}-\nu_{\rm GR})/2} \,, 
\end{equation} 
where $z_{\rm mod} \approx z_{\rm eq}$ for $k> k_{\rm cham}(t_{\rm eq})$ 
and is given by eq.~(\ref{grow}) for 
$k< k_{\rm cham}(t_{\rm eq})$.

In CDM cosmologies structure forms hierarchically (large halos form 
via the merger and accretion of smaller subhalos) and at least some 
substructure is expected to survive. There must be some cut-off in 
this process however; if the density perturbation spectrum extended 
down to infinitely small scales the contribution of density 
perturbations to the local energy density would diverge~\cite{hss}. 
For weakly interacting massive particles (WIMPs) damping 
processes~\cite{damp,hss,bde,wimp1,wimp2,lz}, namely collisional 
damping (due to elastic interactions with radiation) and 
free-streaming, produce a cut-off in the (processed) power  
spectrum at $k_{\rm cut} \sim 1 \, {\rm pc}^{-1}$ (i.e. fluctuations on 
scales $k>k_{\rm cut}$ are erased). Ref.~\cite{wimp2} found a range of 
values $k_{cut} \approx 0.4- 4 \, {\rm pc}^{-1}$ for benchmark models 
spanning the range of plausible WIMP properties. Dirac like WIMPs, 
where elastic scattering is mediated by $Z^{0}$ exchange, 
have values of $k_{\rm cut}$ at the lower end of the range. 
Majorana WIMPs for which $Z^{0}$ exchange is suppressed, for instance 
neutralinos, have a wider range of $k_{\rm cut}$ values 
with more massive WIMPs having larger 
$k_{\rm cut}$. 
 
 \begin{figure}[!ht] 
\label{kchamfig} 
\epsfxsize=12cm 
\begin{center} 
\leavevmode 
\epsffile{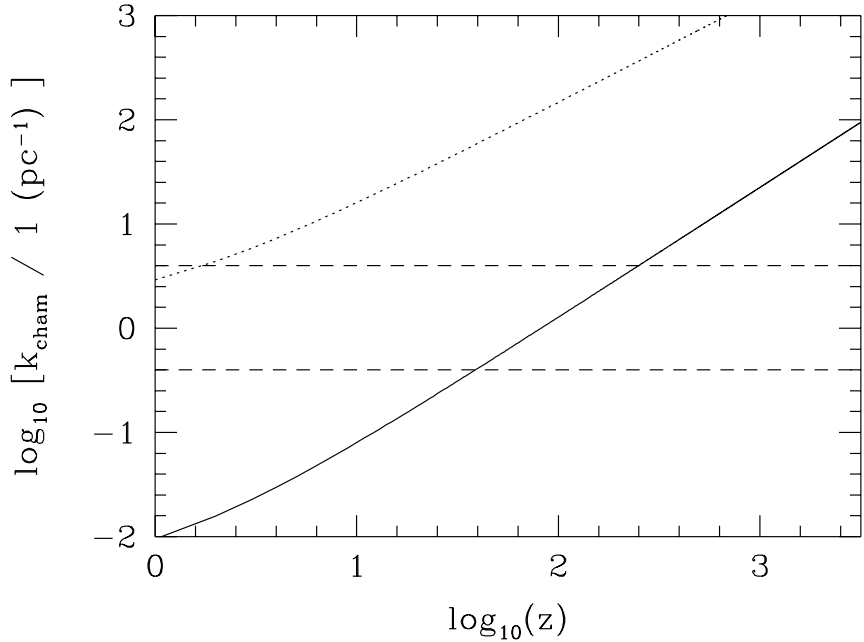} 
\end{center} 
\caption{The red-shift dependence of the 
characteristic wavenumber of the chameleon, 
$k_{\rm cham}(z)$, for $\beta=1$, $M=10^{-3} {\rm eV}$ 
and $n=1$ (solid line) 
and $n=2$ (dotted line). The dashed lines show the range of values of 
$k_{\rm cut}$ for plausible WIMP models from Ref.~\cite{wimp2}.} 
\end{figure}

We plot $k_{\rm cham}(z)$ and $z_{\rm mod}(k)$ for $n=1$ and $n=2$ and 
also the range of $k_{\rm cut}$ values in figs.~1 and 
2 respectively.  For $n=1$ $k_{\rm cham}(z_{\rm eq})$ is 
larger than the upper end of the range of $k_{\rm cut}$ values and the 
evolution of the surviving perturbations is initially unaffected. $k_{\rm 
cham}(z)$ decreases sufficiently rapidly with decreasing red-shift, 
however, that $z_{\rm mod}(k_{\rm cut}) \gg 0$ for the entire range of 
$k_{\rm cut}$ values and the growth law of small (physical) 
scales is modified. For $n=2$ $k_{\rm cham}(z_{\rm eq})$ is so large 
that the chameleon scale is beyond the cut-off scale even at late 
times and the growth of surviving perturbations is completely 
unaffected by the chameleon.  We should emphasise, however, that 
WIMPs are not the only viable CDM candidate. There are a large number 
of other candidates~\cite{dmrev} (including the arguably equally well 
motivated axion) whose microphysics have not yet been studied and 
which may have substantially different cut-off scales.

\begin{figure}[!ht] 
\label{zmodfig} 
\epsfxsize=12cm 
\begin{center} 
\leavevmode 
\epsffile{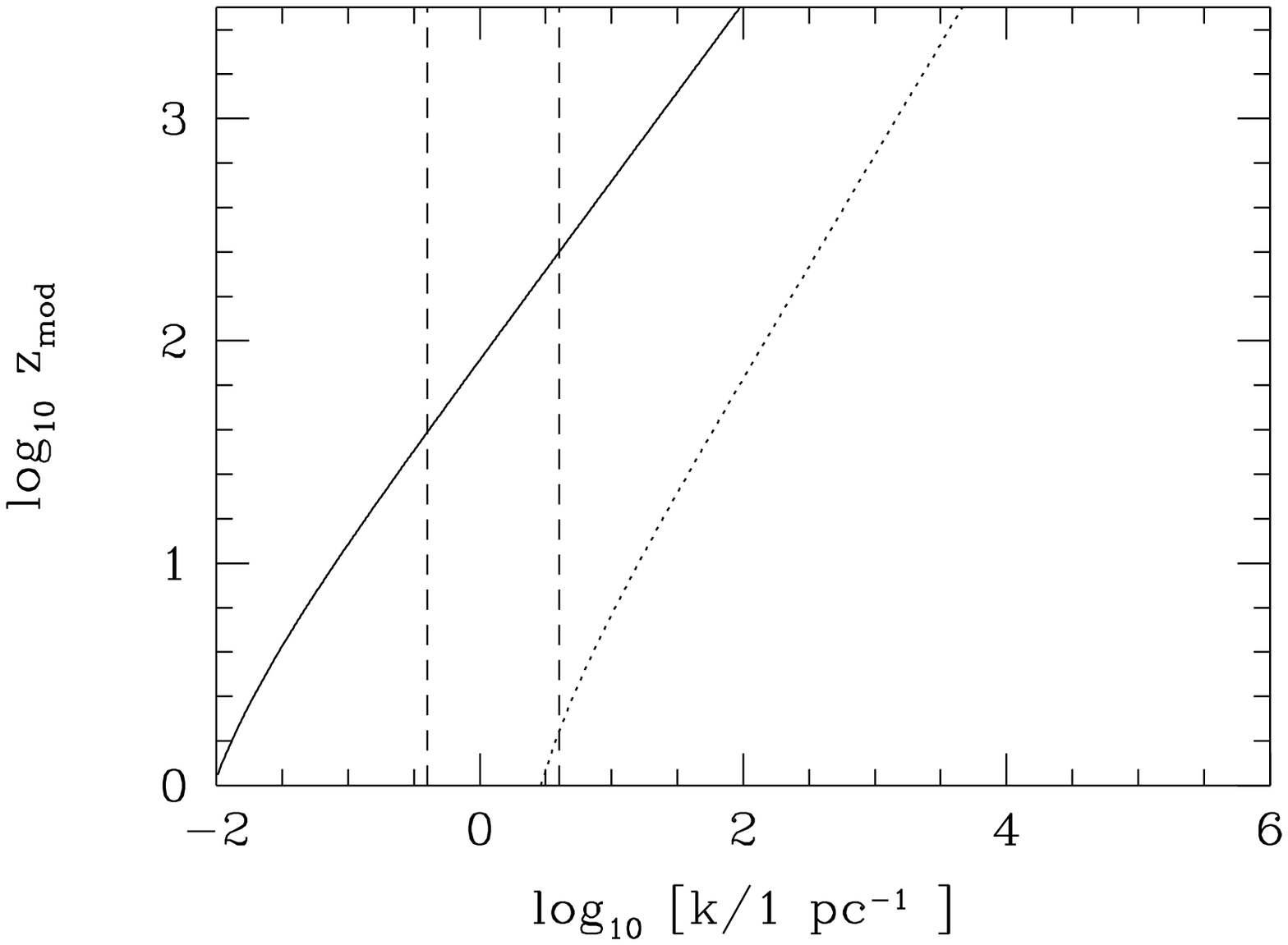} 
\end{center} 
\caption{The red-shift at which the growth law of density
is modified due to the chameleon,
$z_{\rm mod}(k)$, for $\beta=1$, $M=10^{-3} {\rm eV}$ 
and $n=1$ (solid line) 
and $n=2$ (dotted line). The short-dashed lines 
show the plausible range of $k_{\rm cut}$ values for WIMPs. } 
\end{figure}

\section{Small scale structure} 
\label{sscham} 
 
\begin{figure}[!ht] 
\label{znlfig} 
\epsfxsize=12cm 
\begin{center} 
\leavevmode 
\epsffile{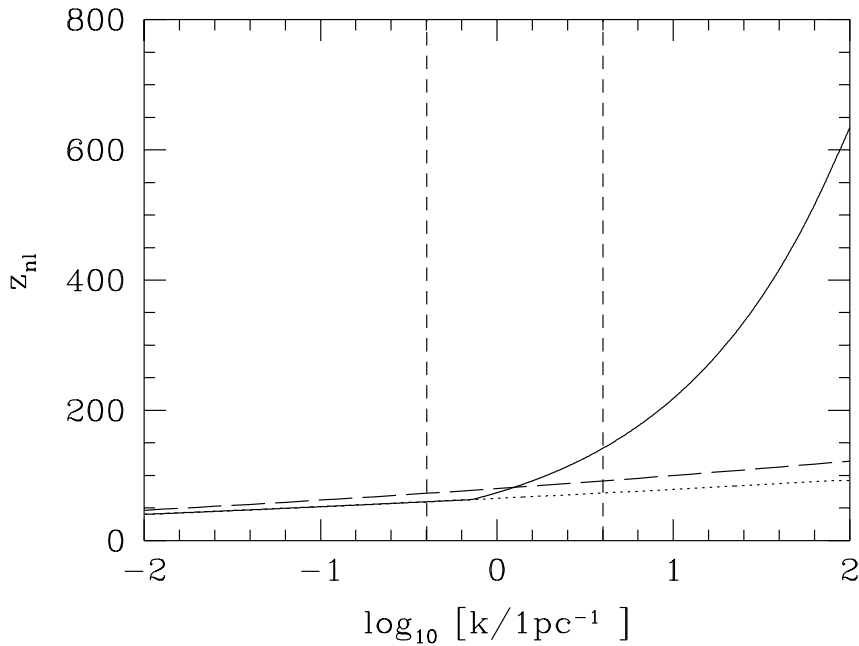} 
\end{center} 
\caption{The red-shift at which typical 1-$\sigma$ fluctuations
collapse to form structure,
$z_{\rm nl,cham}(k)$, for $\beta=1$, $M=10^{-3} {\rm eV}$ and $n=1$ 
(solid line). 
 The dotted line is $z_{\rm nl, GR}(R=1/k)$ and the short-dashed lines 
show the plausible range of $k_{\rm cut}$ values for WIMPs. 
The long-dashed line shows $z_{\rm nl, GR}(R=1/k)$ for a 
primordial density perturbation power spectrum with spectral index  
$n_{\rm s}=1.036$.} 
\end{figure}

The enhancement of the growth rate of the CDM density contrast on 
small scales due to the chameleon means that these scales will go 
non-linear, $\delta_{\rm c} \sim {\cal O}(1)$, and collapse to form 
structure earlier than in standard cosmology.  The red-shift at 
which the first typical, 1-$\sigma$, fluctuations on comoving scale 
$R$ go non-linear $z_{\rm nl}(R)$ is defined via $\sigma(R,z_{\rm 
nl})=1$, where $\sigma(R,z)$ is the mass variance calculated by 
integrating the density perturbation power spectrum multiplied by 
the Fourier transform of a window function. This calculation is not 
possible in this case as we only have the evolution of density 
perturbations on sub-galactic scales, and the change in the growth 
law due to the chameleon means that we can not use the value of 
$\sigma_{8}$ measured by WMAP to evade this problem, as in 
Ref.~\cite{wimp1,wimp2}. We instead estimate the effect of the 
chameleon on the red-shift at which scale go non-linear  by using 
the approximations $\delta_{\rm cham}(k, z_{\rm nl, cham}(k)) = 1$ 
and $\delta_{\rm GR}(k, z_{\rm nl, GR}(k)) = 1$ and using the values 
of $z_{\rm nl, GR}(R)$ calculated in Ref.~\cite{wimp2} with the 
approximation $k \sim 1/R$. Using eq.~(\ref{chamgrow1}) we find 
\begin{equation} 
\frac{1+z_{\rm nl,cham}(k)}{1+z_{\rm nl, GR}(k)}  = 
      \left( \frac{ 1 + z_{\rm mod}(k)}{1+z_{\rm nl, GR}(k)} 
   \right)^{[1-(\nu_{\rm gr}/\nu_{\rm cham})]}  \,. 
\end{equation} 
The resulting values of $z_{\rm nl, cham}(k)$ are plotted in 
Fig.~\ref{znlfig} For scales with $k< k_{\rm cham}(t_{\rm eq})$ we 
assume that $\nu$, and hence the growth law for $\delta_{\rm c}$, 
changes abruptly at $z_{\rm mod}$. The change would in fact occur 
smoothly as the chameleon term in square brackets in eq.~(\ref{del1}) 
increases smoothly from 1 to ($1 + 2 \beta^2$). This approximation is 
reasonable however given the other uncertainties involved in the 
calculation.

A scale dependent primordial power spectrum, with spectral index $n_{\rm s} > 
1$, as produced by, for instance, false vacuum dominated hybrid 
inflation would also result in a larger than standard density contrast 
on small physical scales and hence earlier structure formation. We 
therefore also plot $z_{\rm nl, GR}$ as calculated in 
Ref.~\cite{wimp2} for a false vacuum dominated hybrid inflation model 
which produces a primordial power spectrum with $n_{\rm s}=1.036$, which is 
the maximum scale dependence allowed by the WMAP and 2dF 
data~\cite{ll}\footnote{A possible interaction between the chameleon and  
the inflaton field does not change the result for $n_s$, since  
the chameleon field is rather heavy during inflation and has  
no influence on the effective mass of the inflaton field \cite{chamus}.}.   
For $n=2$ $z_{\rm nl, cham}(k)=z_{\rm nl, GR}(k)$ for 
all $k< 10^{2} \, {\rm pc}^{-1}$. For $n=1$,  for 
$k > {\cal O}( 1 \, {\rm pc}^{-1})$ the growth law is modified 
sufficiently early that $z_{\rm nl, cham}(k) \gg z_{\rm nl, 
GR}(k)$. The rapid increase of $z_{\rm nl, cham}(k)$ with increasing 
$k$ is caused by the large change in the index of the growth law by 
the chameleon, $\delta_{\rm c} \propto a^{1.9}$ compared with $\delta_{\rm c} 
\propto a$ for standard cosmology. This resulting change in 
$\delta_{\rm c}$ on small physical scales at late times is far larger 
than that produced by a primordial power spectrum with $n_{\rm s}=1.036$. 
 
\begin{figure}[!ht] 
\label{del} 
\epsfxsize=12cm 
\begin{center} 
\leavevmode 
\epsffile{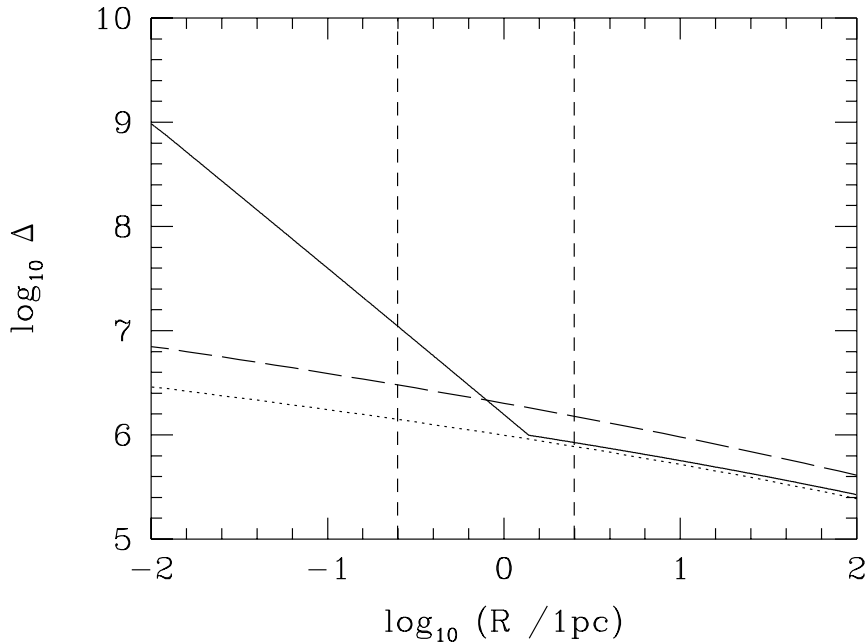} 
\end{center} 
\caption{The present day overdensity corresponding to typical fluctuations,
$\Delta$, for the chameleon with 
$\beta=1$, $M=10^{-3} {\rm eV}$ and $n=1$ 
(solid line) and for standard cosmology with 
primordial power spectra which are scale independent (dotted) 
and $n_{\rm s}=1.036$ (long-dashed). The short-dashed lines 
show the range of $1/k_{\rm cut}$ values for plausible WIMPs.} 
\end{figure}

The physical properties of the first generation of WIMP halos  were 
estimated in Ref.~\cite{wimp2} using the spherical collapse model. 
The physical size of the halos post collapse is proportional to 
$(1+z_{\rm nl})^{-1}$ and hence the density contrast is proportional 
to $(1+z_{\rm nl})^{3}$ i.e. the earlier halos form the more 
over-dense they are at later times (reflecting the higher matter 
density at the time they form). In fig.~\ref{del} we plot the 
present day overdensity corresponding to
 typical halos, which form from 1-$\sigma$ 
fluctuations, 
 $\Delta$,  
\begin{equation} 
\Delta = \frac{2 M(R)}{\frac{4 \pi}{3} r(R)^3 
   \rho_{\rm m}(t_{0})} \,,
\end{equation} 
as a function of comoving scale
where $M(R)= 1.6 \times 10^{7} M_{\odot} (\Omega_{\rm m} h^2/ 0.14) 
(R/{\rm pc})^3$ is the mean mass within a sphere of comoving radius $R$ 
and $r(R)=0.53 R/(1+z_{\rm nl})$ is the physical, post collapse, radius of a 
halo which forms from  a typical fluctuation with comoving size 
$R$ at red-shift 
$z_{\rm nl}$. For the chameleon we make the approximation $R \sim 1/k$. 
The first halos to 
form in chameleon cosmology are significantly more concentrated than 
in standard cosmology (provided $k_{\rm cut} > {\cal O} (1 {\rm 
pc}^{-1})$ if the CDM is in the form of WIMPs) 
and are hence more likely to resist disruption by dynamical 
processes (such as tidal disruption and interactions with stars). We 
therefore expect that the present day dark matter distribution on 
small scales would be more clumped than in standard cosmology and this 
could be detectable via axion detectors or WIMP direct and indirect 
detection experiments (the survival 
probability of the first dark matter halos even in standard cosmology 
is the subject of ongoing studies~\cite{bde,dms,othersmall}). 
This alone would not provide a `smoking gun' 
for the chameleon, as other modifications of general relativity could 
also lead to enhanced growth of small scale density perturbations and 
hence small scale structure. However, combined with a detection of 
modified gravity within the solar system~\cite{exp} the present day 
density distribution could be used to probe modifications of gravity, 
such as those due to the chameleon.

{\bf Acknowledgements:} This work was supported in part by PPARC.

\end{document}